\documentstyle[aps,preprint]{revtex}

\begin{document}
\draft
\title{Antitrace maps and light transmission coefficients for a generalized Fibonacci multilayers}
\author{Xiaoguang Wang \quad Shaohua Pan  \quad Guozhen Yang}
\address{Laboratory of Optical Physics, Institute of Physics, 
Chinese Academy of Sciences, \\
Beijing 100080}

\date{\today}
\maketitle

\begin{abstract}
By using antitrace map method, we investigate the light transmission for a generalized Fibonacci multilayers.
Analytical results are obtained for transmission coefficients in some special cases.
We find that the transmission coefficients possess two-cycle property or six-cycle property.  
The cycle properties of the trace and antitrace are also obtained.
\end{abstract}


The transmission of light through the multilayers arranged by the Fibonacci\cite{Optics,Jin96,Huang99}, non-Fibonacci
sequence\cite{Dulea90,Riklund88}, Thue-Morse sequence\cite{Liu97}, and the generalized Thue-Morse sequence\cite{Kolar91} was
studied in the
literature. Schwartz\cite{Schwartz88}
suggested a possibility of quasiperiodic multilayers as optical switches and memories. Huang {\it et al}\cite{Huang93} and
Yang {\it et al}\cite {XBYang99} have found an interesting switch-like property in the light transmission
through Fibonacci-class sequences.

On the other hand, the trace-map technique, first introduced in 1983\cite{Kohmoto83}, has
proven to be a powerful tool to investigate the properties of
various aperiodic systems. However, as pointed by Dulea {\it et al}\cite{Dulea90},
we must know the so-called ``antitrace map'' when we evaluate the light transmission
coefficients through aperoidic sequences. 
 Here, the so-called ``antitrace" of a
$2\!\times\! 2$ matrix
\begin{equation}
A = \left(\matrix{A_{11} & A_{12}\cr
                  A_{21} & A_{22}}\right)
\end{equation}
is defined as $y=A_{21}-A_{12}$.
Recently we have  
given a detailed study on the antitrace maps for various aperiodic systems and shown that the antitrace maps
exist
for arbitrary substitution sequences\cite{WangGrimm}.

In this paper, we use the antritrace map method to evaluate the transmission coefficients through a generalized Fibonacci
sequence. Its substitution rule is\cite{GFS}
\begin{equation}
b\rightarrow a, \quad a\rightarrow a^nb.
\end{equation}

Now we
consider that the light transmit vertically through the generalized Fibonacci 
multilayer which is sandwiched by two media of type $a$.  The corresponding transfer
matrices $A_l$ are written as\cite{Optics}
\begin{eqnarray}
A_1&=&P_{ab}{P}_bP_{ba},\nonumber\\
A_2&=&{P}_a,\nonumber\\
A_{l+1}&=&A_l^nA_{l-1},\label{eq:mat}
\end{eqnarray}
where ${P}_{ab} ({P}_{ba})$ stands for the propagation matrix from
layer $a (b)$ to $b (a)$ and ${P}_{a} ({P}_b) $ is the propagation
matrix through single layer $a (b)$. They are given by\cite{Optics}
\begin{eqnarray}
P_{ab}&=&P_{ba}^{-1}=\left(\matrix{1&0\cr 0&n_a/n_{b}}\right),\nonumber\\
P_{a(b)}&=&\left(\matrix{\cos\delta_{a(b)}&-\sin\delta_{a(b)}\cr 
\sin\delta_{a(b)}&\cos\delta_{a(b)}}\right),\label{eq:matmat}
\end{eqnarray} 
where $\delta_{a(b)}=kn_{a(b)}d_{a(b)}$, $n_{a(b)}$ is the refraction
index of media ${a(b)}$, $d_{a(b)}$ are the thickness of layers, and
$k$ is the wave number in vacuum.  The quantity $\delta_{a(b)}$ is the
phase difference between the ends of a layer.

We remark here that the trace and antitrace map of the generalized Fibonacci sequence is identical to 
that of the Fibonacci-class sequence\cite{XBYang99} since they have the same recursion relations for 
the transfer matrix. However the initial two transfer matrices $A_1$ and $A_2$ are different. 

The transmission coefficient is given by\cite{Optics}
\begin{equation}
T_l=\frac{4}{|A_l|^2+2},
\end{equation}
where $|A_l|^2$ is the sum of squares of the four elements of
$A_l$. Since the transfer matrix is unimodular, we can
express the transmission coefficient in the following form
\begin{equation}
T_l=\frac{4}{x_l^2+y_l^2},
\label{eq:t}
\end{equation}
where $x_l$ and $y_l$ denote the trace and antitrace of the transfer
matrix $A_l$, respectively. 

{}From Eq.~(\ref{eq:t}), we see that the transmission coefficient is
completely determined by the trace and antitrace, i.e., a complete
description of the light transmission through general aperiodic
multilayers requires both the trace and antitrace map\cite{Dulea90}.

In the following discussion we need to know the $n$th power of a unimodular $2\!\times\! 2$ matrix $A$ , which can be
written as\cite{Kolar90,Baake,Wang,WangPan}
\begin{equation}
A^n=U_n(x_A)A-U_{n-1}(x_A)I,
\label{eq:an}
\end{equation} 
where $I$ is the unit matrix and 
\begin{eqnarray}
U_n(x_A)&=&\frac{\lambda_+^n-\lambda_-^n}{\lambda_+-\lambda_-},\nonumber\\
\lambda_\pm&=&\frac{x_A\pm\sqrt{x_A^2-4}}{2}.
\end{eqnarray}
Here $x_A$ and $\lambda_\pm$ denote the trace and the two
eigenvalues of $A$, respectively. 

Using Eq.(\ref{eq:an}), we can write the recursion relation of the transfer
matrix (\ref{eq:mat}) as
\begin{equation}
A_{l+1}=U_n(x_l)A_lA_{l-1}-U_{n-1}(x_l)A_{l-1}.\label{eq:eq9}
\end{equation}

From the above equation, the trace map is easily obtained as
\begin{eqnarray}
x_{l+1}&=&U_n(x_{l})v_l-U_{n-1}(x_{l})x_{l-1},\nonumber\\
v_{l+1}&=&U_{n+1}(x_{l})v_l-U_{n}(x_{l})x_{l-1},\label{eq:trace}
\end{eqnarray}
where $v_{l}=\text{tr}(A_lA_{l-1})$ is a subsidiary quantity.

In order to study antitrace maps we need the following identity
for two unimodular transfer matrices $A$ and $B$
\cite{Dulea90}
\begin{equation}
y_{AB}=x_By_A+x_Ay_B-y_{BA},
\label{eq:yab1}
\end{equation}
where $y_A$ denotes the antitrace.

Using the above equantion and Eq.~(\ref{eq:eq9}), we obtain the antitrace map as
\begin{eqnarray}
y_{l+1}&=&U_n(x_{l})\bar{w}_l-U_{n-1}(x_{l})y_{l-1},\nonumber\\
\bar{w}_{l+1}&=&x_{l+1}y_l+U_{n-1}(x_{l})\bar{w}_l-U_{n-2}(x_{l})y_{l-1}.
\label{eq:antitrace}
\end{eqnarray}
Here $\bar{w}=y_{A_{l}A_{l-1}}$ is also subsidiary.
The trace and antitrace map are completely determined by
Eqs.~(\ref{eq:trace}) and (\ref{eq:antitrace}). The forms of trace and
antitrace maps are different from that in Refs.\onlinecite{Dulea90} and
\onlinecite{XBYang99}. These forms are easy to be obtained and are convenient for
application. If we know the initial conditions, the transmission
coefficients can be determined from the trace and antitrace map. 
Note that the coefficients in Eq.(\ref{eq:antitrace}) are dependent on the traces of
the transfer matrices. We must know the trace map when we make use of 
the antitrace map.

We choose appropriate thickness of the
layers $d_a$ and $d_b$ to make $n_ad_a=n_bd_b$. Then we have
$\delta_a=\delta_b=\delta$.
For $\delta=(k+1/2)\pi$, the propagation matrices $P_{a(b)}$ become
\begin{eqnarray}
{P}_{a}&=&{P}_{b}=\left(\matrix {0&-1\cr 1&0} \right).\label{eq:aaa}
\end{eqnarray}

Then from Eqs.(\ref{eq:mat}), (\ref{eq:matmat}), and  (\ref{eq:aaa})
the initial conditions for
the
trace and antitrace map are obtained as
\begin{eqnarray}
x_1&=&0,\nonumber\\
x_2&=&0,\nonumber\\
v_2&=&-(R+R^{-1}),\nonumber\\
y_1&=&R+R^{-1},\nonumber\\
y_2&=&2,\nonumber\\
\bar{w}_2&=&0,
\end{eqnarray}
where $R=n_a/n_b$. 
The initial conditions depend on a single parameter $R$.

Now we classify the generalized Fibonacci sequence into two classes, i.e, the even family with $n=2m$ and the odd family with
$n=2m+1$, m=0,1,2,....

First we consider the even family.
From the trace map, the initial conditions and the following property of the function $U_n(x)$
\begin{eqnarray}
U_{2m}(0)&=&0, \; U_{2m+1}(0)=(-1)^m,\label{eq:eq15}
\end{eqnarray}
we see that the trace $x_l=0$ for any $l$. Therefore the transmission coefficient is only dependent on the antitrace $y_l$.
From the antitrace map equation and the initial conditions, we obtain the antitrace $y_l$ and the transmission coefficients.
The result is shown in Table I.

\begin{center}
Table I. The antitrace and transmission coefficients for the even family.\\

\vspace{0.5cm}  
\begin{tabular}{|c|c|c|c|}\hline
$ l$ & $y_l(\text{odd}\; m)$ &   $y_l(\text{even}\; m)$ & $T_l$ \\ \hline 
$1$ &   $R+R^{-1}$  &   $R+R^{-1}$  & $\frac{4}{R^2+R^{-2}+2}$\\ \hline
$2$ &   2&   2&   1\\ \hline
$3$ &   $-(R+R^{-1})$&   $(R+R^{-1})$ & $\frac{4}{R^2+R^{-2}+2}$ \\ \hline
$4$ &     $-2$   &  2 & 1 \\ \hline
\end{tabular}
\end{center}

We see the antitrace has four-cycle property (odd $m$) or two-cycle property (even $m$). 
The transmission coefficients have two-cycle property. For even $l$, the transmission
coefficient is 1, while for odd $l$, the transmission coefficient is ${4}/({R^2+R^{-2}+2})$.
Actually from the first line of Eq.(\ref{eq:antitrace}) and Eq.~(\ref{eq:eq15}), we obtain
\begin{equation}
y_{l+1}=(-1)^my_{l-1}.
\end{equation} 
This leads to Table I immediately.

Next we consider the odd family.
From the trace map, antitrace map, and the initial conditions, we obtain Table II.

\begin{center}
Table  II. The antitrace and transmission coefficients for the odd family.
The upper sign refers to even, and the lower sign refers to odd values of $m$.

\vspace{0.5cm}
\begin{tabular}{|c|c|c|c|}\hline
$l $  &$x_l$& $y_l$ & $T_l$ \\  \hline  
$1$ &0                &   $R+R^{-1}$          &       $\frac{4}{R^2+R^{-2}+2}$\\ \hline
$2$ &0                &   2                   &       1\\ \hline
$3$ &$\mp(R+R^{-1})$  &0                      &       $\frac{4}{R^2+R^{-2}+2}$ \\ \hline
$4$ &0                & $\mp(R^n+R^{-n})$          &       $\frac{4}{R^{2n}+R^{-2n}+2}$ \\ \hline
$5$ &0                & $\pm(R^{n-1}+R^{-n+1})$   &       $\frac{4}{R^{2n-2}+R^{-2n+2}+2}$ \\ \hline
$6$ &$\pm(R+R^{-1})$  & 0                     &       $\frac{4}{R^2+R^{-2}+2}$ \\ \hline
\end{tabular}
\end{center}

The trace $x_l$, antitrace $y_l$ and the transmission coefficients all have six-cycle property.
From Table II, we know that the trace and antitrace is not completely same for odd and even $m$.
They have a sign difference. The trace and antitrace are zero alternatively.
We also see that the the transmission coefficient is 1 for $l=6k+2 (k=0,1,2,...)$, i.e., the light is transparent
in this case.

In conclusion, we have studied the light transmission for the generalized Fibonacci sequences by using
the antitrace method. The analytical results for the transmission coefficient are obtained in some special cases. 
The transmission coefficients have two-cycle property for even $n$ and six-cycle property 
for odd $n$. The analysis in this paper shows that it is very convenient to use the antitrace map method to evaluate the transmission coefficients.
This method can be used not only in the generalized Finonacci sequences, but also in other substitution sequences.

\acknowledgments

The author Xiaoguang Wang thanks for many useful discussions with Dr. Uwe Grimm, Technische Universitaet Chemnitz, Germany.

\end{document}